\documentclass[a4paper]{article}

\usepackage{makecell}
\usepackage{xcolor}
\usepackage{url}
\usepackage{nicefrac}
\usepackage{booktabs}
\usepackage{multirow}
\usepackage{INTERSPEECH2019}
\usepackage{amssymb}
\usepackage{pifont}

\newcommand{\cmark}{\ding{51}}%
\newcommand{\xmark}{\ding{55}}%
\newcommand{\newpara}[1]{\vspace{5pt}\noindent\textbf{#1}}
\usepackage[implicit=false]{hyperref}

\title{The ins and outs of speaker recognition: lessons from VoxSRC 2020}
\name{Yoohwan Kwon$^{1,2}$*, Hee-Soo Heo$^{1}$*, Bong-Jin Lee$^{1}$, Joon Son Chung$^{1}$\thanks{\hspace{-12pt}* These authors contributed equally to this work.}}
\address{$^1$Naver Corporation, $^2$Yonsei University
\email{}}

\begin{document}

\maketitle
\begin{abstract}
The VoxCeleb Speaker Recognition Challenge (VoxSRC) at Interspeech 2020 offers a  challenging evaluation for speaker recognition systems, which includes celebrities playing different parts in movies.
The goal of this work is robust speaker recognition of utterances recorded in these challenging environments. 
We utilise variants of the popular ResNet architecture for speaker recognition and perform extensive experiments using a range of loss functions and training parameters. 
To this end, we optimise an efficient training framework that allows powerful models to be trained with limited time and resources.
Our trained models demonstrate improvements over most existing works with lighter models and a simple pipeline.
The paper shares the lessons learned from our participation in the challenge.

\end{abstract}


\section{Introduction}

Speaker recognition has various applications such as personalisation of voice command systems and user authentication in security systems. 
Such applications require speaker recognition systems to be robust to performance degradation in real-world scenarios.
For example, noise and channel effects have significant effect on the performance of speaker recognition systems.
In order to address these issues, the VoxCeleb dataset has been released, allowing researchers to tackle the problem of speaker recognition ``in the wild''~\cite{Nagrani17, Chung18a}. 
The research community in this field has quickly adopted the dataset and has made remarkable technological advances. 
For example, some studies have directly dealt with issues such as channel difference and noise~\cite{luu2020channel,chung2020delving}, while others have incorporated advances in deep learning architectures and loss functions for improving speaker recognition performance~\cite{snyder2018x,Jung2018AResult,ravanelli2018speaker,okabe2018attentive,snyder2019speaker, wan2018generalized}.

Fast progress of the field have revealed the need for a new challenging dataset, since the performance has saturated as a result of powerful models overcoming most of the known errors. 
While the best performance on the original VoxCeleb test set was reported to be 7.8\% (equal error rate) in 2017~\cite{Nagrani17} and 3.8\% in 2018~\cite{okabe2018attentive}, some recent works have demonstrated results under 1\%~\cite{desplanques2020ecapa}.

The VoxCeleb Speaker Recognition Challenge 2020 identifies the limitations in existing test datasets and proposes new challenges for the speaker recognition community. 
This year's challenge is different to the last in a number of ways: (1) there is an explicit domain shift between the training data and the test data; (2) the dataset contains celebrities playing different characters in movies, providing more challenging target trials; (3) the test set contains utterances that are shorter than the segments seen during training. 

Our submission to the challenge is based on popular architectures such as the Residual Networks (ResNet) and Time Delay Neural Networks (TDNN)~\cite{he2016deep, waibel1989phoneme, snyder2018x}. We perform a wide range of experiments using our efficient training framework, and discuss strengths and weaknesses of the different loss functions. 

\section{Methods}

\subsection{Trunk architectures}

We experiment with three different trunk architectures -- two variants of the ResNet-34 used in~\cite{heo2020clova} and one TDNN-based model introduced in~\cite{desplanques2020ecapa}.

\newpara{Speed optimised ResNet.} 
Residual networks~\cite{he2016deep} are widely used in image recognition and have been successfully applied to speaker recognition~\cite{Chung18a,cai2018exploring,Xie19a,chung2020delving,kwon2020intra}. 
In order to reduce computational cost, the speed optimised variant uses one {\bf quarter} of the channels in each residual block compared to the original ResNet-34. The model has 1.4 million parameters compared to 22 million of the original ResNet-34. Self-attentive pooling (SAP)~\cite{cai2018exploring} is used to aggregate frame-level features into utterance-level representation, while paying attention to the frames that are more informative for utterance-level speaker recognition. The network architecture is identical to that used in~\cite{chung2020defence} except for the dimension of the input features, and we refer to this configuration as {\bf Q~/~SAP} in the results.

\newpara{Performance optimised ResNet.} 
The performance optimised variant has {\bf half} of the channels in each residual block compared to the original ResNet-34, and in total contains 8.0 million parameters. The stride at the first convolutional layer is removed, leading to increased computational requirement compared to the speed optimised model. Attentive Statistics Pooling (ASP)~\cite{okabe2018attentive} is used to aggregate temporal frames, where the channel-wise weighted standard deviation is calculated and concatenated to the weighted mean. 
Table~\ref{table:model} shows the detailed architecture of this model, referred to as {\bf H~/~ASP} in the results.

\newpara{ECAPA-TDNN.}
ECAPA-TDNN~\cite{desplanques2020ecapa} consists of a series of 1-dimensional Res2Net layers followed by statistics pooling module with channel-dependent frame attention. There are two variants proposed in the paper, but we use the larger model with 14.7 million parameters. Our implementation of the model is identical to that in the original paper, but differ in pre and post-processing steps used during training and evaluation. This is the best performing single model on the VoxCeleb1 test dataset in the recent literature.

We train all three models with and without batch normalisation after the linear embedding layer.

\begin{table}[h]

 \caption{Trunk architecture for the performance optimized ResNet model. {$\mathbf{L}$}: length of input sequence in frames, {\bf ASP}: attentive statistics pooling.}
  \centering
  \vspace{-5pt}
  \label{table:model}
  \begin{tabular}{l c c c}
  
  \toprule

   Layer & Kernel size  & Stride& Output shape   \\
  \midrule
  Conv1 & $3 \times 3 \times 32$ & $1\times1$& $L \times 64\times 32$ \\
  \hline
  Res1 & $3\times 3 \times 32$ & $1\times1$ & $ L \times64 \times 32$ \\
  \hline
  Res2 & $3\times 3\times 64$ &$2\times2$ & $ \nicefrac{L}{2} \times32 \times 64$ \\
  \hline
  Res3 & $3\times 3 \times 128$ &$2\times2$ & $ \nicefrac{L}{4} \times16 \times 128$\\
  \hline
  Res4 & $3\times 3\times 256$ &$2\times2$ & $ \nicefrac{L}{8} \times8 \times 256$\\
  \hline
  Flatten & - & - & $ \nicefrac{L}{8} \times 2048$\\
  \hline
  ASP & - & - & $ 4096$ \\
  \hline
  Linear & $ 512 $ & - & $ 512 $ \\
  \bottomrule
  \end{tabular}
\vspace{-10pt}
\end{table}
\begin{table*}[!t]
\caption{Results on the VoxCeleb and VoxSRC test sets.
The figures in bold represent the best results for each model and metric.
{\bf AP:} Angular Prototypical.
{\bf Aug.:} Data augmentation used during training.
{\bf BN:} Batch normalisation used after the embedding layer.
$\dag$: This method uses score normalisation as a post-processing step.
$\ddag$: Additional data augmentation methods are applied.
$\S$: Submissions to VoxSRC 2019.}
\label{table:results}
\scriptsize
\setlength{\tabcolsep}{4pt}
\renewcommand\arraystretch{1.2}
\vspace{-5pt}
\centering
\begin{tabular}{ r l l l l l | r r | r r | r r | r r | r r  }
\toprule
 & \textbf{Config.} & \textbf{\# Params} & \textbf{Loss} & \textbf{Aug.} & \textbf{BN}   & \multicolumn{2} {c|} {\bf VoxCeleb1 cl.} & \multicolumn{2} {c|} {\bf VoxCeleb1-E cl.} & \multicolumn{2} {c|} {\bf VoxCeleb1-H cl.} & \multicolumn{2} {c|} {\bf VoxSRC 2019} & \multicolumn{2} {c} {\bf VoxSRC 2020 Val} \\
   & &     & &  & & \textbf{EER} & \textbf{MinDCF} & \textbf{EER}  & \textbf{MinDCF}  & \textbf{EER} & \textbf{MinDCF}  & \textbf{EER} & \textbf{MinDCF} & \textbf{EER} & \textbf{MinDCF} \\ 
\midrule

& FR-34~\cite{chung2020defence}& 1.4M & AP           & \xmark & \xmark  & 2.05 & 0.166 & 2.27 & 0.164 & 4.36 & 0.282 & 2.65 & 0.182           & 6.35 & 0.374 \\
& Sys 1~\cite{zeinali2019but} $\S$ & 6.1M & Softmax~$\dag$ & \cmark & \xmark  & 1.42 & - & 1.35 & - & 2.48 & - & - & -        & - & - \\
& Fusion~\cite{zeinali2019but} $\S$ & -  & - & - & -  & 1.02 & - & 1.14 & - & 2.21 & - & 1.42 & -        & - & - \\

& Sys A5~\cite{garcia2020jhu} $\S$ & 60M & AM-Softmax & \cmark & \xmark & - & -  & 1.51 & - & - & - & 1.72 & -        & - & - \\
& Fusion~\cite{garcia2020jhu} $\S$ & - &  - & - & - & - & - & 1.22 & - & - & - & 1.54 & -        & - & - 
\\
& ECAPA 1K~\cite{desplanques2020ecapa} & 6.2M & AAM-Softmax~$\dag$ & \cmark~$\ddag$ & \cmark & 1.01 & -  & 1.24 & - & 2.32 & - & 1.32 & -        & - & - \\
& ECAPA 1K~\cite{desplanques2020ecapa} & 14.7M & AAM-Softmax~$\dag$ & \cmark~$\ddag$ & \cmark & 0.87 & -  & 1.12 & - & 2.12 & - & 1.22 & -        & - & - 
\\\midrule\midrule
Q1  & \multirow{8}{*}{Q / SAP} & \multirow{8}{*}{1.4M} & AM-softmax  & \xmark & \xmark & 1.67 & 0.121 & 1.96 & 0.131 & 3.54 & 0.213 & 2.20 & 0.132 & 5.22 & 0.295 \\
Q2  &          &  & AAM-softmax & \xmark & \xmark & 1.57 & 0.112 & 1.75 & 0.118 & 3.24 & 0.199 & 2.09 & 0.124 & 4.82 & 0.270 \\
Q3  &          &  & AP          & \xmark & \xmark & 1.58 & 0.138 & 1.87 & 0.136 & 3.62 & 0.239 & 2.14 & 0.150 & 5.32 & 0.320 \\
Q4  &          &  & AP+softmax  & \xmark & \xmark & 1.47 & 0.119 & 1.78 & 0.130 & 3.44 & 0.229 & 2.04 & 0.143 & 5.09 & 0.310 \\
Q5  &          &  & AM-softmax  & \xmark & \cmark & 1.60 & 0.128 & 1.80 & 0.119 & 3.20 & 0.194 & 2.17 & \textbf{0.123} & 4.79 & 0.268 \\
Q6  &          &  & AAM-softmax & \xmark & \cmark & 1.50 & \textbf{0.110} & 1.75 & \textbf{0.117} & \textbf{3.09} & \textbf{0.190} & 2.00 & \textbf{0.123} & \textbf{4.68} & \textbf{0.260} \\
Q7  &          &  & AP          & \xmark & \cmark & 1.51 & 0.115 & 1.72 & 0.125 & 3.40 & 0.224 & 1.98 & 0.140 & 5.10 & 0.298 \\
Q8  &          &  & AP+softmax  & \xmark & \cmark & \textbf{1.45} & 0.111 & \textbf{1.68} & 0.119 & 3.21 & 0.206 & \textbf{1.99} & 0.129 & 4.85 & 0.277 \\ \midrule
QA1 & \multirow{8}{*}{Q / SAP}  & \multirow{8}{*}{1.4M} & AM-softmax  & \cmark & \xmark & 1.64 & \textbf{0.105} & 1.88 & 0.124 & 3.36 & 0.206 & 2.19 & 0.131 & 5.05 & 0.280 \\
QA2 &          &  & AAM-softmax & \cmark & \xmark & 1.65 & 0.123 & 1.77 & 0.119 & 3.19 & 0.199 & 2.02 & 0.128 & 4.87 & 0.272 \\
QA3 &          &  & AP          & \cmark & \xmark & 1.69 & 0.131 & 2.06 & 0.147 & 4.00 & 0.257 & 2.46 & 0.160 & 5.93 & 0.339 \\
QA4 &          &  & AP+softmax  & \cmark & \xmark & 1.51 & 0.129 & 1.67 & 0.123 & 3.28 & 0.214 & 2.01 & 0.135 & 5.03 & 0.294 \\
QA5 &          &  & AM-softmax  & \cmark & \cmark & 1.71 & 0.115 & 1.90 & 0.122 & 3.33 & 0.200 & 2.17 & 0.130 & 5.03 & 0.276 \\
QA6 &          &  & AAM-softmax & \cmark & \cmark & 1.60 & 0.112 & 1.77 & 0.120 & 3.19 & 0.197 & 2.08 & 0.128 & 4.84 & \textbf{0.267} \\
QA7 &          &  & AP          & \cmark & \cmark & 1.54 & 0.124 & 1.69 & 0.125 & 3.37 & 0.222 & 2.05 & 0.139 & 5.14 & 0.303 \\
QA8 &          &  & AP+softmax  & \cmark & \cmark & \textbf{1.37} & 0.116 & \textbf{1.59} & \textbf{0.112} & \textbf{2.98} & \textbf{0.191} & \textbf{1.86} & \textbf{0.123} & \textbf{4.63} & \textbf{0.267} \\ \midrule
H1  & \multirow{8}{*}{H / ASP}  & \multirow{8}{*}{8.0M} & AM-softmax  & \xmark & \xmark & 1.77 & 0.128 & 2.00 & 0.130 & 3.51 & 0.208 & 2.20 & 0.130 & 5.12 & 0.288 \\
H2  &          &  & AAM-softmax & \xmark & \xmark & 1.74 & 0.132 & 1.85 & 0.125 & 3.30 & 0.199 & 2.10 & 0.121 & 4.74 & 0.267 \\
H3  &          &  & AP          & \xmark & \xmark & 1.44 & 0.116 & 1.74 & 0.122 & 3.35 & 0.211 & 1.92 & 0.131 & 4.95 & 0.287 \\
H4  &          &  & AP+softmax  & \xmark & \xmark & \textbf{1.21} & \textbf{0.098} & 1.42 & \textbf{0.099} & \textbf{2.77} & 0.175 & 1.64 & \textbf{0.101} & 4.15 & 0.241 \\
H5  &          &  & AM-softmax  & \xmark & \cmark & 1.44 & 0.101 & 1.62 & 0.107 & 2.85 & 0.174 & 1.84 & 0.111 & 4.28 & 0.243 \\
H6  &          &  & AAM-softmax & \xmark & \cmark & 1.25 & 0.101 & 1.56 & 0.103 & 2.78 & \textbf{0.166} & 1.75 & 0.105 & \textbf{4.12} & \textbf{0.231} \\
H7  &          &  & AP          & \xmark & \cmark & 1.20 & 0.102 & 1.52 & 0.110 & 3.01 & 0.194 & 1.70 & 0.116 & 4.50 & 0.270 \\
H8  &          &  & AP+softmax  & \xmark & \cmark & 1.29 & 0.091 & \textbf{1.41} & 0.100 & 2.78 & 0.176 & \textbf{1.55} & 0.109 & 4.21 & 0.241 \\ \midrule
HA1 & \multirow{8}{*}{H / ASP}  & \multirow{8}{*}{8.0M} & AM-softmax  & \cmark & \xmark & 1.53 & 0.113 & 1.67 & 0.114 & 3.07 & 0.191 & 1.89 & 0.112 & 4.59 & 0.263 \\
HA2 &          &  & AAM-softmax & \cmark & \xmark & 1.49 & 0.111 & 1.51 & 0.105 & 2.91 & 0.181 & 1.75 & 0.103 & 4.40 & 0.245 \\
HA3 &          &  & AP          & \cmark & \xmark & 1.24 & 0.105 & 1.50 & 0.108 & 3.06 & 0.198 & 1.66 & 0.112 & 4.60 & 0.267 \\
HA4 &          &  & AP+softmax  & \cmark & \xmark & \textbf{0.88} & \textbf{0.079} & \textbf{1.07} & \textbf{0.076} & \textbf{2.21} & \textbf{0.147} & \textbf{1.26} & \textbf{0.079} & \textbf{3.51} & \textbf{0.202} \\
HA5 &          &  & AM-softmax  & \cmark & \cmark & 1.35 & 0.101 & 1.45 & 0.099 & 2.64 & 0.166 & 1.64 & 0.097 & 4.05 & 0.232 \\
HA6 &          &  & AAM-softmax & \cmark & \cmark & 1.15 & 0.083 & 1.35 & 0.091 & 2.49 & 0.155 & 1.61 & 0.093 & 3.80 & 0.218 \\
HA7 &          &  & AP          & \cmark & \cmark & 1.16 & 0.086 & 1.38 & 0.099 & 2.78 & 0.180 & 1.62 & 0.103 & 4.32 & 0.248 \\
HA8 &          &  & AP+softmax  & \cmark & \cmark & 1.03 & 0.084 & 1.23 & 0.087 & 2.47 & 0.159 & 1.40 & 0.091 & 3.94 & 0.220 \\ \midrule

EA1 & \multirow{8}{*}{ECAPA 1K} & \multirow{8}{*}{14.7M} & AM-softmax  & \cmark & \xmark & 1.13 & 0.085 & 1.35 & 0.093 & 2.61 & 0.165 & 1.54 & 0.090 & 4.03 & 0.228 \\
EA2 &          &  & AAM-softmax & \cmark & \xmark & 0.96 & 0.076 & 1.24 & 0.086 & 2.40 & \textbf{0.152} & 1.59 & 0.085 & 3.82 & \textbf{0.212} \\
EA3 &          &  & AP          & \cmark & \xmark & 1.14 & 0.088 & 1.40 & 0.100 & 2.88 & 0.189 & 1.72 & 0.105 & 4.41 & 0.258 \\
EA4 &          &  & AP+softmax  & \cmark & \xmark & \textbf{0.90} & 0.081 & \textbf{1.11} & \textbf{0.077} & \textbf{2.32} & 0.155 & \textbf{1.32} & 0.084 & \textbf{3.81} & 0.219 \\
EA5 &          &  & AM-softmax  & \cmark & \cmark & 1.23 & 0.093 & 1.33 & 0.091 & 2.52 & 0.157 & 1.54 & 0.093 & 3.85 & 0.221 \\
EA6 &          &  & AAM-softmax & \cmark & \cmark & 1.01 & \textbf{0.077} & 1.27 & 0.085 & 2.45 & 0.156 & 1.64 & \textbf{0.082} & 3.83 & \textbf{0.212} \\
EA7 &          &  & AP          & \cmark & \cmark & 1.12 & 0.084 & 1.35 & 0.098 & 2.78 & 0.183 & 1.51 & 0.103 & 4.37 & 0.250 \\
EA8 &          &  & AP+softmax  & \cmark & \cmark & 0.96 & 0.078 & 1.22 & 0.083 & 2.45 & 0.163 & 1.49 & 0.090 & 4.00 & 0.228 \\ \midrule
F1    & \multirow{2}{*}{Fusion}   & - & Best EER    &             -          &      -                 & 0.73 & 0.056 & 0.93 & 0.065 & 1.87 & 0.122 & 1.18 & 0.066 & 3.08 & 0.174 \\
F2    &          &  - & Best MinDCF &           -            &             -          & 0.74 & 0.061 & 0.93 & 0.066 & 1.90 & 0.124 & 1.17 & 0.067 & 3.10 & 0.173
\\ \bottomrule
\end{tabular} 
\end{table*}

\subsection{Loss function}

Additive margin softmax (AM-softmax)~\cite{wang2018additive,wang2018cosface} and Additive angular margin softmax (AAM-softmax)~\cite{deng2019arcface} introduce a concept of margin between classes in order to increase inter-class variance.
For AM-Softmax and AAM-Softmax losses, we select a margin of 0.2 and a scale of 30, since these values give the best results on the VoxCeleb1 test set.

The Angular Prototypical (AP) loss~\cite{chung2020defence} is a variant of the prototypical networks~\cite{snell2017prototypical} that uses an angular objective, and has demonstrated strong performance without the need for manual hyper-parameters tuning~\cite{chung2020defence,huh2020augmentation}. Our implementation uses $M=2$, with one utterance for every speaker in the query set and one utterance in the support set.

\begin{figure}[!htb]
\centering 
\includegraphics[width=0.7\columnwidth]{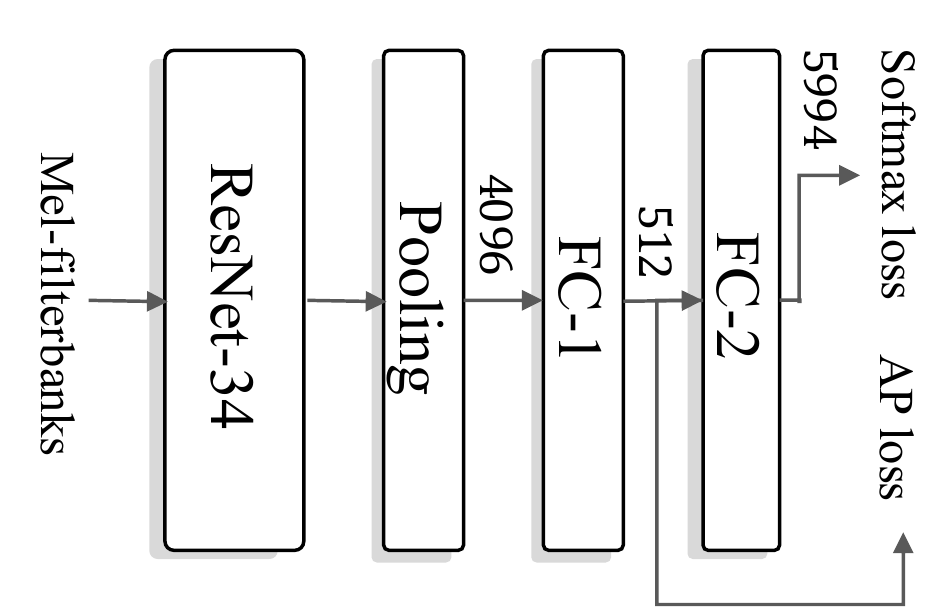}
\vspace{-5pt}
\caption{Overview of the architecture and the configuration of the AP+Softmax loss.}
\label{fig:overview} 
\end{figure}

Finally, we combine the Angular Prototypical loss together with the vanilla softmax loss which provide further improvement in performance over using each of the loss functions alone. Figure~\ref{fig:overview} shows the training strategy for using the AP and the softmax losses jointly. The use of multiple loss functions has been discussed in~\cite{kye2020meta,Heo2019}.

\section{Experiments}

\subsection{Dataset}

The models are trained on the development set of VoxCeleb2~\cite{Chung18a}, which contains 5,994 speakers. The VoxCeleb1 test sets~\cite{Nagrani17}, the previous year's VoxSRC test set~\cite{chung2019voxsrc} and this year's VoxSRC development set are used for validation.

\subsection{Input representations}

During training, we use random 2-second temporal segments extracted from each utterance. Pre-emphasis with a coefficient of 0.97 is applied to the input signal. The spectrograms are extracted with a hamming window of 25ms width and 10ms step size.

For the ResNet models, the input features are 64-dimensional log Mel-filterbanks.
Mean and variance normalisation (MVN) is performed by applying instance normalisation~\cite{ulyanov2016instance} to the input.

For the TDNN models, the input features are 80-dimensional Mel-frequency cepstral coefficients (MFCCs) extracted from the spectrograms. The features are mean normalised at the input to the network.

We do not use voice activity detection during training and testing, since active speaker detection in the data collection pipeline of the VoxCeleb dataset removes any non-speech segments.

\subsection{Data augmentation}

Augmentation increases the amount and diversity of the training data, which helps reduce overfitting.
We employ two of the popular augmentation methods in speech processing -- additive noise and room impulse response (RIR) simulation. For additive noise, we use the audio clips from the MUSAN corpus~\cite{snyder2015musan}; for RIR, we sample the simulated filters of small and medium rooms released in~\cite{ko2017study}.
  The augmentation parameters are selected to replicate the behaviour of the Kaldi recipe of~\cite{snyder2018x}.

In contrast to the previous works that utilise the Kaldi implementation for data augmentation~\cite{snyder2018x,desplanques2020ecapa}, we implement an efficient implementation of the augmentation methods so that they can be preformed online in the data loader. This allows the models to be trained in machines without large amounts of local storage, since the augmented version of the dataset do not need to be stored. Moreover, this allows different noises and RIR filters to be applied at every epoch, and therefore allows the model to observe unlimited variations of the utterances during training. Tempo augmentation was not applied because the waveform similarity overlap-add (WSOLA) algorithm used by the previous works is computationally expensive and only provides a very marginal improvement of performance.


\subsection{Implementation details}

Our implementation is adapted from the PyTorch-based speaker recognition code released with~\cite{chung2020defence}.

The models are trained using NVIDIA V100 GPUs with 32GB memory using the Adam optimiser, with an initial learning rate of 0.001. The batch size of 200 is used for the classification-based loss functions and 400 for the metric learning-based losses. A weight decay of {\em 5e-5} is applied. Training schedules vary by the model as smaller models take more epochs to converge.

We use mixed precision training, a technique to train deep neural networks using half precision floating point numbers. This allows larger batch sizes to be used for any given GPU memory size, which can significantly boost the performance of metric learning-based models~\cite{chung2020defence}.

\newpara{Speed optimised ResNet.} 
The learning rate is reduced by $5\%$ every 5 epochs. The network is trained for 300 epochs.

\newpara{Performance optimised ResNet.} 
The learning rate is reduced by $25\%$ every 16 epochs. The network is trained for 250 epochs.

\newpara{ECAPA-TDNN.} 
The learning rate is reduced by $25\%$ every 8 epochs. The network is trained for 150 epochs.

\subsection{Scoring}

The trained networks are validated on the VoxCeleb and the VoxSRC test sets. We sample ten 4-second temporal segments at regular intervals from every segment in the test set, and compute the $10 \times 10 = 100$ similarities for every pair using all possible combinations of segments. The mean of the 100 similarities is used as the final pairwise score. This protocol is referred to as test time augmentation (TTA-3) in~\cite{Chung18a} and is used by~\cite{chung2020delving,chung2020defence}.

\subsection{Evaluation protocol}

We use two performance metrics: (i) the Equal Error Rate (EER) which corresponds to the threshold at which the false acceptance rate is equal to the false rejection rate; and (ii) the minimum detection cost (MinDCF) of the function used by the NIST SRE~\cite{nist2018}
and the VoxSRC\footnote{\url{http://www.robots.ox.ac.uk/~vgg/data/voxceleb/competition2020.html}} evaluations. 
The parameters $C_{miss}=1$, $C_{fa}=1$ and $P_{target}=0.05$ are used for the cost function.

\subsection{Fusion}

We perform the feature fusion by computing the weighted average of the score of individual systems.
We search over all possible combinations of models and weights in order to compensate for the difference of the back-end model.
The weight value is chosen from 0, 1, 2 or 3 for each model.
We optimise the fusion weights on the VoxSRC 2020 validation set in order to find the optimal parameters that give the best EER or MinDCF.
For the best model based on the EER, the weights of 3 were given to {\bf HA4} and {\bf EA4}, while {\bf EA5} and {\bf HA6} got weights of 1.

\subsection{Results}

Table~\ref{table:results} compares the results on the validation sets. 
We compare our models to two of the best scoring submissions~\cite{zeinali2019but,garcia2020jhu} to VoxSRC 2019.

We select the models that perform best on the validation sets, and evaluate them on VoxSRC 2020 test set. The results on the test set are reported in Table~\ref{table:testres}.

\begin{table}[h]
 \caption{Results on the challenge test set}
  \centering
  \vspace{-5pt}
  \label{table:testres}
  \begin{tabular}{r | c c}
  \toprule
  {\bf Model}  & \multicolumn{2} {c} {\bf VoxSRC 2020} \\
   &  {\bf EER} & {\bf MinDCF} \\ \midrule
  HA4 & 4.98 & 0.294 \\
  F1 & 4.22 & 0.250 \\
  \bottomrule
  \end{tabular}
\end{table}

\subsection{Discussion}

\begin{figure}[!htb]
\centering 
\includegraphics[width=1\columnwidth]{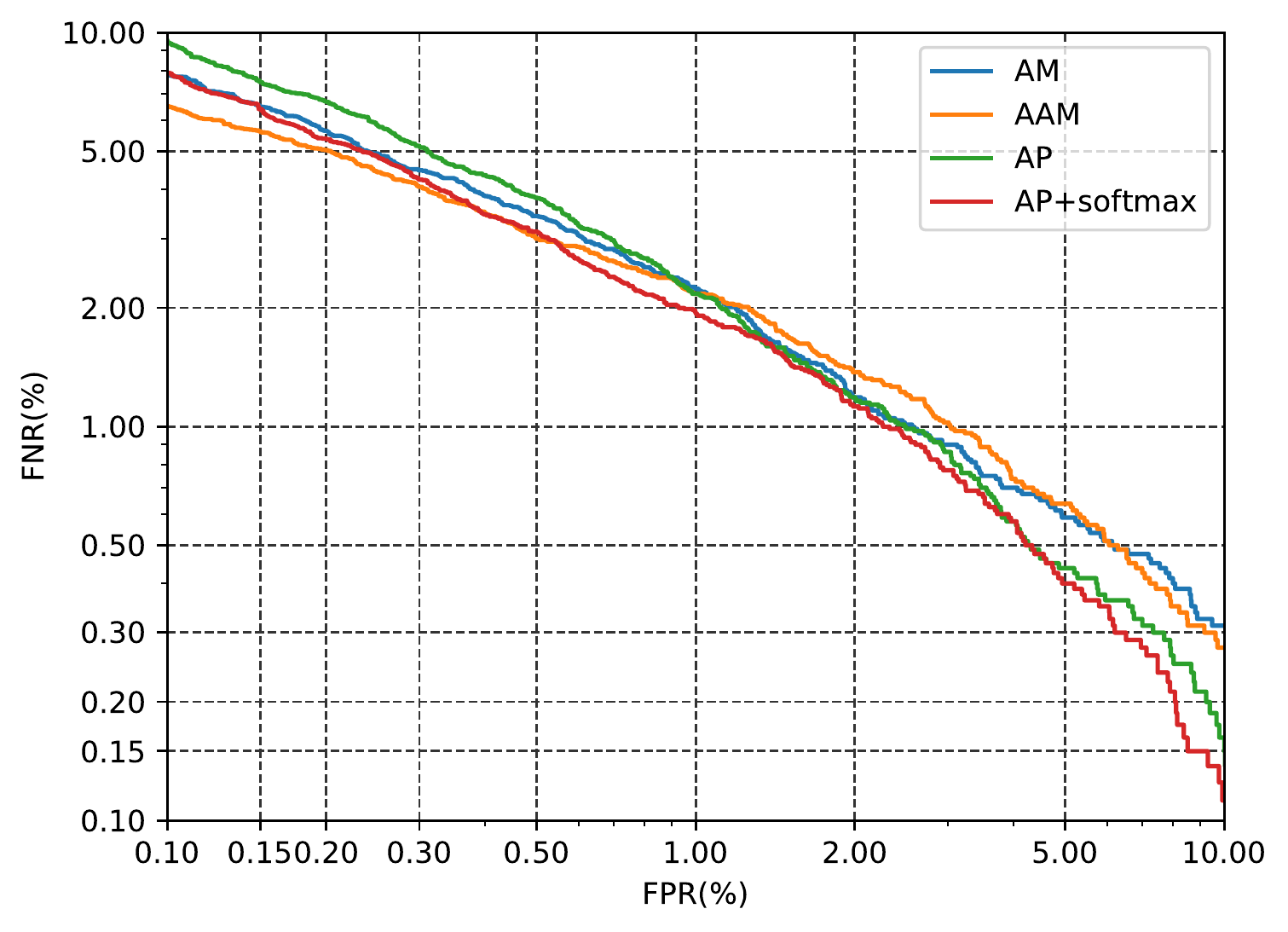}
\vspace{-10pt}
\caption{DET curves for ECAPA-TDNN models with batch normalisation on the output.}
\label{fig:det} 
\end{figure}

The results show that the combination of metric learning and classification objectives work best for most models. The batch normalisation applied after the embedding layer contributes a significant performance improvements to the models trained with AM-softmax and AAM-softmax losses. 

\newpara{Analysis of the DET curve.}
Figure~\ref{fig:det} represents the detection error trade-off (DET) curve for the ECAPA-TDNN models trained with batch normalisation using various loss functions. It can be seen that the model trained with AP+Softmax loss shows the best performance for most operating points, but AAM-softmax trained model works best at operating points where false positives carry a high penalty. The DET curves for the other model types also follow similar patterns -- AP-based models are generally well suited to low FNR scenarios ({\em e.g.} personalisation of music preference for AI speakers), whereas AM-Softmax and AAM-Softmax work well for low FPR scenarios ({\em e.g.} telephone banking authentication).

\newpara{Effects of data augmentation.}
We perform experiments with and without data augmentation for both speed optimised and performance optimised ResNet models. Data augmentation contributed a significant boost in performance for variants of the larger performance optimised model, which is in line with expectations. However, in the case of the smaller speed optimised model, data augmentation had adverse effect on performance for almost all variants, despite the fact that the model was left to train for a long time until convergence. Therefore it can be concluded that data augmentation only helps to improve the performance when the model has enough capacity to take advantage of the variations introduced by the augmentation methods.

\newpara{Effects of batch size.} The experiments in~\cite{chung2020defence} have shown that the performance of models trained with metric learning objectives increases with the batch size. We verify the phenomenon for the {\bf H~/~ASP} model trained with AP and AP+Softmax losses in Table~\ref{table:batch}. Note that our implementation of the model and that of~\cite{heo2020clova} are identical except for the larger batch size, enabled by the use of more modern GPUs and mixed precision training.

\begin{table}[h]
 \caption{Comparison of the results for different batch sizes. The {\bf H~/~ASP} trunk architecture is used. {\bf BS}: Batch size.}
  \centering
  \vspace{-5pt}
  \label{table:batch}
  \begin{tabular}{l l l l | c c}
  \toprule
  {\bf BS} & {\bf Loss} & {\bf Aug.} & {\bf BN} & \multicolumn{2} {c} {\bf VoxSRC 2019} \\
  & &  & & {\bf EER} & {\bf MinDCF} \\ \midrule
 150~\cite{heo2020clova} & AP        & \cmark & \xmark & 1.92 & 0.128 \\
 150~\cite{heo2020clova} & AP+Softmax & \cmark & \xmark & 1.46 & 0.088 \\
 150~\cite{heo2020clova} & AP        & \cmark & \cmark & 1.74 & 0.117 \\
 150~\cite{heo2020clova} & AP+Softmax & \cmark & \cmark & 1.49 & 0.102 \\ \midrule
 400 & AP        & \cmark & \xmark & 1.66 & 0.112 \\
 400 & AP+Softmax & \cmark & \xmark & 1.26 & 0.079 \\
 400 & AP        & \cmark & \cmark & 1.62 & 0.103 \\
 400 & AP+Softmax & \cmark & \cmark & 1.40 & 0.091 \\
  \bottomrule
  \end{tabular}
\end{table}

\newpara{Comparison to~\cite{desplanques2020ecapa}.} Our best model ({\bf HA4}) trained with the AP+Softmax loss matches the performance of the large ECAPA-TDNN using a model that has only a half the number of parameters, and with a more simple implementation. Our systems do not use tempo augmentation, re-encoding or SpecAugment during training. Moreover, we do not use score normalisation as a post-processing step on the output. As a result of these differences, the models trained with our implementation under-performs compared to the results reported in~\cite{desplanques2020ecapa} for the same model specification and loss function.

\section{Conclusion}

The paper describes our experience from the participation in VoxSRC 2020. The best system is trained using a combination of metric learning and classification-based objectives, and outperform the baselines by a significant margin. Our best model outperforms all single model and ensemble systems submitted to the last year's challenge.

\vspace{3pt}
\newpara{Acknowledgements.}
We would like to thank Brecht Desplanques for his help with the implementation of ECAPA-TDNN. 

\clearpage
\raggedbottom
\bibliographystyle{IEEEbib}
\bibliography{shortstrings,mybib}
\end{document}